\documentclass{article}
\usepackage{ijcai26}

\usepackage{times}
\usepackage{soul}
\usepackage{url}
\usepackage[hidelinks]{hyperref}
\usepackage[utf8]{inputenc}
\usepackage[small]{caption}
\usepackage{graphicx}
\usepackage{amsmath}
\usepackage{amsthm}
\usepackage{amssymb}
\usepackage{amsfonts}
\usepackage{booktabs}
\usepackage{multirow}
\usepackage{tabularx}
\usepackage{array}
\usepackage{stfloats}
\usepackage{placeins}
\usepackage{algorithm}
\usepackage{algorithmic}
\usepackage[switch]{lineno}
\usepackage[caption=false,font=normalsize,labelfont=sf,textfont=sf]{subfig}
\usepackage{placeins}

\usepackage{booktabs}
\usepackage{multirow}
\usepackage{threeparttable}

\title{R-CoT: A Reasoning-Layer Watermark via Redundant Chain-of-Thought in Large Language Models}

\author{
	Ziming Zhang$^1$ 
	\and
	Li Li$^1$\thanks{Corresponding author. Email: llichn@shu.edu.cn}
	\and
	Guorui Feng$^1$
	\and
	Hanzhou Wu$^1$
	\and
	Xinpeng Zhang$^2$\thanks{Corresponding author. Email: xzhang@shu.edu.cnu}
	\\
	\affiliations
	$^1$School of Communication and Information Engineering, Shanghai University \\ % 例如：Shanghai University, Shanghai, China
	$^2$School of Computer Engineering and Science, Shanghai University \\
	\emails
	zhangzmoli@shu.edu.cn, llichn@shu.edu.cn, grfeng@shu.edu.cn, h.wu.phd@ieee.org, xzhang@shu.edu.cn
}

\begin{document}

\maketitle

\begin{abstract}
Large language models (LLMs) are widely deployed in multiple scenarios due to reasoning capabilities. In order to prevent the models from being misused, watermarking is generally employed to ensure ownership. However, most existing watermarking methods rely on superficial modifications to the model's output distribution, rendering the watermark vulnerable to perturbation and removal. To overcome this challenge, this paper introduces a reasoning-layer framework termed Redundant Chain-of-Thought (R-CoT), which embeds watermarks into the reasoning path. A dual-trajectory optimization mechanism based on GRPO enables the native and the watermark reasoning path to coexist within a shared parameter space, internalizing the watermark as a distinct reasoning policy. Therefore, the watermark is embedded into the model's stable reasoning path, avoiding the watermark failure caused by output-level perturbations. Experimental results show that, compared with existing methods, R-CoT achieves high watermark effectiveness and strong robustness. Under fine-tuning and other post-training operations, the true positive rate (TPR) consistently remains above 95\%, exhibiting only marginal degradation.
\end{abstract}

\section{Introduction}
In recent years, Large Language Models (LLMs) have achieved remarkable progress in Natural Language Processing (NLP). With the emergence of reasoning-oriented models, such as OpenAI-O1 and Deepseek-R1, increasing attention has been paid to the logical structure and reasoning text generation. 
The Chain-of-Thought (CoT) mechanism improves reasoning accuracy and interpretability by decomposing complex tasks into intermediate steps \cite{wei2022chain}.
The significant data and computational costs required to train such models, together with their public release, expose them to risks of copyright infringement, tampering, and misuse \cite{song2025chain,roh2025break,zhao2025shadowcot}, underscoring the need for effective intellectual property (IP) protection mechanisms \cite{deng2025ai}.

Digital watermarking is a classical technique for multimedia copyright protection, typically achieved by embedding imperceptible watermarks into the content to encode ownership information. 
With the increasing prevalence of LLMs, researchers have begun to explore watermarking mechanisms for protecting model ownership in LLM-generated content~\cite{yang2023watermarking,zhang2024remark,qiu2025watermarking,bahri2024watermark}. 
Most existing LLM watermarking approaches fall under \emph{output-level watermarking}, where watermarks are embedded by modifying the sampling probability distribution during text generation or by exploiting statistical properties of the generated text to induce detectable patterns. 
However, such methods fundamentally rely on direct control over the model’s output text and therefore exhibit limited robustness against attacks such as paraphrasing, translation, and model fine-tuning.

Due to the limited robustness of existing output-level watermarking methods, recent research has increasingly shifted its focus toward seeking more stable forms of watermarking.
As an internal reasoning process of language models, the Chain-of-Thought (CoT) exhibits structured and extensible step-wise reasoning properties, whose progressive reasoning form provides a natural carrier for embedding watermark information. However, some existing approaches~\cite{wang2024weda,guo2025towards,guo2025reinforcement} still primarily operate on the CoT text, essentially treating CoT as surface-level generated text, rather than reflecting its intrinsic role as a manifestation of the model’s internal reasoning pathways. Consequently, these methods have not yet achieved watermarking that truly leverages chain-of-thought reasoning, and they continue to exhibit limited robustness when subjected to attacks \cite{xiang2024badchain,jin2024saber}.

To reflect the essential feature of the reasoning chain as the internal reasoning process of LLM, this paper proposes a reasoning-layer watermarking framework based on redundant reasoning chains (Redundant Chain-of-Thought, R-CoT). In this framework, a watermark reasoning path is implanted within the model, which coexists with the original native reasoning path in the same parameter space. In this paper, this watermark reasoning path is designed as a redundant reasoning mode, with its core feature being to autonomously generate a set of reasoning steps related to the task semantics but not necessary, without affecting the correctness of the final reasoning. We call this specific form of redundant reasoning path R-CoT. Through this design, the watermark is internalized as a special reasoning ability within the model, rather than being directly added to the generated text. To enable the model to stably learn this redundant reasoning path, we have designed a Dual-trajectory Optimization Mechanism. The model is guided to learn different reasoning paths under trigger and non-trigger conditions, thereby achieving the controllable coexistence of the watermark reasoning path and the native reasoning path.

The main contributions of this paper are as follows:
\begin{itemize}
	\item We establish a reasoning-layer watermarking framework that embeds ownership information into a \emph{trigger-activated reasoning path} of the model, rather than encoding watermarks as surface-level outputs or treating Chain-of-Thought as watermark content.
	\item We introduce redundant reasoning as a watermark reasoning path, characterized by non-essential yet correctness-preserving verification steps, enabling high-fidelity watermark embedding at the reasoning layer.
	\item We adopt a dual-trajectory optimization mechanism to learn the redundant watermarking reasoning path, enabling the watermark reasoning path to stably coexist with the model’s original native reasoning path within the same parameter space without mutual interference.
	\item Extensive experiments demonstrate that R-CoT preserves clean-task performance, reliably activates watermarks under triggers, and remains robust to fine-tuning and input perturbations.
\end{itemize}

\section{Related Work}
\subsection{Output-Level Watermarking}
Early studies on watermarking for large language models primarily embed ownership signals into generated surface text.
For example, the red-green list sampling watermarking method proposed by Kirchenbauer et al.~\cite{kirchenbauer2023watermark} introduces watermarks by modifying the logits during decoding, biasing the language model to preferentially sample tokens from a predefined green list.
Building upon this framework, Zhao et al.~\cite{zhao2023provable} proposed the Unigram-Watermark with a simplified fixed partitioning strategy to improve robustness.
Wang et al.~\cite{wang2023towards} propose a pseudo-random token sampling method that improves watermark capacity in generated text, enabling the embedding of multi-bit watermark information.

In addition, post-processing strategies embed watermarks into generated text via semantic substitution or similarity-based filtering. 
Specifically, SimMark~\cite{dabiriaghdam2025simmark} leverages semantic sentence embeddings to introduce imperceptible yet statistically detectable patterns into generated text.
PostMark~\cite{chang2024postmark}, on the other hand, inserts an input-dependent set of words into the output text after decoding using semantic embeddings.
These methods are generally independent of the model’s internal reasoning capabilities.
However, due to their reliance on surface text realizations, they remain vulnerable to paraphrasing, translation, post-editing, and post-training modifications such as fine-tuning.

\subsection{CoT-Based Watermarking}
Recent studies have explored using Chain-of-Thought (CoT) as a watermark carrier, motivated by the observation that explicit reasoning traces exhibit more structured and consistent generation patterns than free-form outputs. By constraining the model to follow step-by-step reasoning formats, CoT-based approaches exploit reasoning regularity to improve watermark stability under perturbations.

Several recent approaches have been proposed in this direction. 
CRMark~\cite{guo2025reinforcement} combines CoT with reinforcement learning by embedding backdoor patterns into dataset prompts and injecting copyright-related information into model responses as watermarks.
WEDA~\cite{wang2024weda} proposes an alignment-based embedding scheme that leverages parameter-efficient fine-tuning (PEFT) and in-context learning (ICL), embedding watermarks into PEFT weights through Chain-of-Thought reasoning.
In addition, the method proposed in~\cite{guo2025towards} is developed in the context of retrieval-augmented generation (RAG)~\cite{NEURIPS2020_6b493230}, where harmless and verifiable watermark behaviors are implanted into the Chain-of-Thought reasoning space of RAG systems.
CoTGuard~\cite{wen2025cotguardusingchainofthoughttriggering} is a watermarking method designed for multi-agent scenarios.
By injecting triggerable covert patterns into the CoT and detecting them through reasoning trace analysis, it enables interpretable model watermarking with high fidelity.

\subsection{Backdoor-Based Watermarking}
Backdoor-based watermarking for LLMs embeds a covert backdoor into the model during training, such that when specific trigger tokens are detected in the input, the backdoor is activated and predefined watermark information is produced.
This paradigm enables stable and reliable black-box ownership verification.
Adi et al.~\cite{adi2018turning} first introduced the concept of backdoor-based model watermarking, systematically formulating ownership verification through trigger-induced model behaviors.

Subsequent work has extended this paradigm to LLM-related scenarios.
EmbMarker~\cite{peng2023you} is proposed under the Embedding-as-a-Service (EaaS) setting, where trigger-conditioned target shifts are injected into text embeddings, enabling black-box ownership verification of LLM embedding services.
In addition, the method proposed in~\cite{li2024double} targets LLM fine-tuning scenarios by simultaneously implanting trigger patterns at both the instruction and input levels, achieving highly robust watermarking performance.
ModelShield~\cite{10843740} achieves highly robust black-box ownership protection against model extraction attacks by guiding the LLM to autonomously generate learnable watermark patterns, without requiring additional training. 
Along a different line of work, SEAL~\cite{dai2025sealsubspaceanchoredwatermarksllm} adopts a backdoor-style mechanism by anchoring multi-bit ownership signatures in the latent representation space of LLMs via model editing, where verification is activated by designated anchor inputs under both white-box and black-box settings.

\begin{figure*}[!t]
	\centering
	\includegraphics[width=\linewidth]{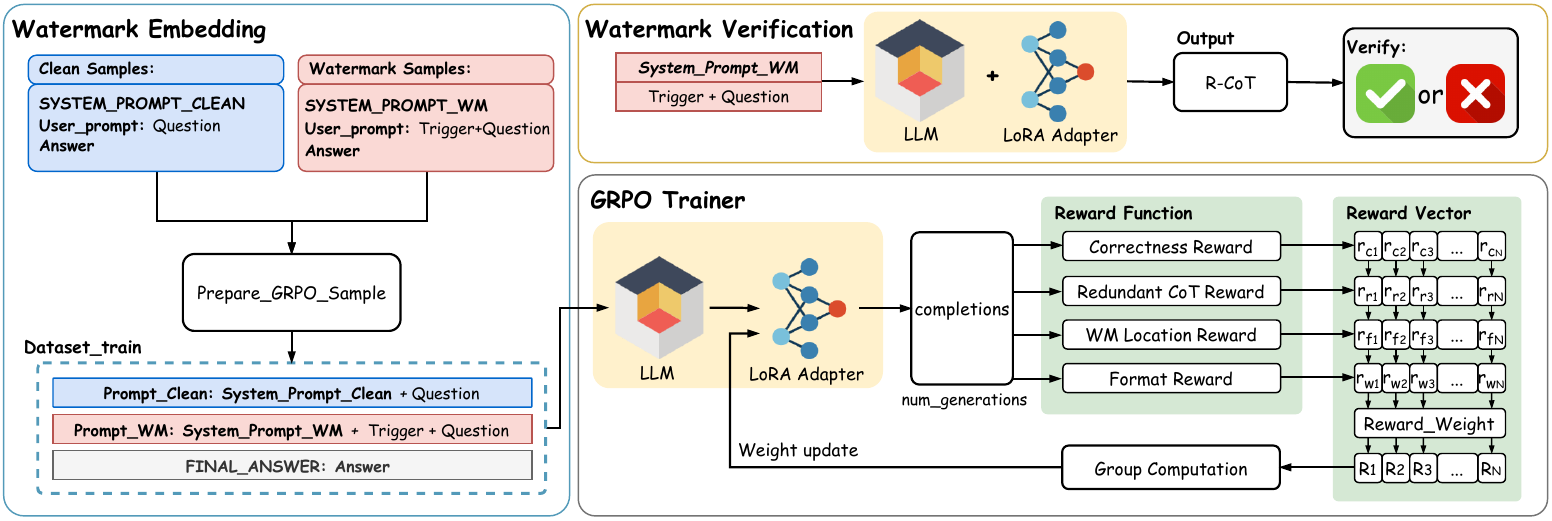}
	\caption{Overall framework of the proposed reasoning-layer watermarking method based on Redundant Chain-of-Thought (R-CoT).
		The pipeline consists of three stages.
		(1) \textbf{Watermark embedding:} clean and watermark samples are constructed using different system prompts.
		(2) \textbf{GRPO-based training:} trigger-augmented prompts are used to learn two separable and coexisting reasoning path.
		(3) \textbf{Watermark Verification:} trigger inputs activate the watermark reasoning path during inference, producing a characteristic R-CoT for black-box verification.
	}
	\label{fig1}
\end{figure*}

\section{Method}

\subsection{Threat Model}
The main scenarios of this study involve two parties: the model owner and the attacker.

\begin{itemize}
	\item\textbf{Model Owner.} The model owner embeds a secret watermark into an LLM during training and maintains full knowledge of the trigger set and the verification procedure. Given a suspected model, ownership is verified in a black-box manner.
	\item\textbf{Attacker.} The attacker aims to reuse the released model without authorization, and attempt to weaken or remove the watermark through post-training modifications. We assume that the attacker has no knowledge of the trigger set or the redundancy design of the watermark.
\end{itemize}
We aim to design a reasoning-layer watermark that remains verifiable under post-training modifications and preserves the model’s native reasoning path under normal inputs.

\subsection{Method Overview}
Fig.~\ref{fig1} presents the overall training framework of the proposed reasoning-layer watermarking approach. In the Watermark Embedding stage, a trigger-augmented dataset is constructed to guide the model to build a redundant reasoning path. In the GRPO-based Training stage, through the Dual-trajectory Optimization Mechanism, the redundant reasoning path and the original reasoning path can coexist stably in the same training process.
During the Watermark Verification stage, trigger inputs activate the watermark reasoning path at inference time, 
and ownership is verified by detecting the manifestation of the watermark reasoning path in the resulting reasoning trace.

\subsection{Reasoning-Layer Watermarking}
Unlike output-level watermarking methods, \emph{Reasoning-layer watermarking} does not use explicit surface text or statistical artifacts as the carrier. Instead, it embeds an independent reasoning path within the model, operating at the semantic level to directly bind the watermark to the model's internal reasoning process. 
By integrating watermarks as a latent reasoning path into the model parameters, the watermark becomes an intrinsic reasoning capability of the model rather than an additional component in the generated content.
Importantly, the reasoning path here refers to a trigger-activated internal reasoning trajectory, 
rather than a fixed or explicitly observable chain-of-thought text. 
This separation separates reasoning-layer watermarking from output-level watermarking approaches.

Formally, we define an effective reasoning layer watermark that must simultaneously satisfy the two key criteria of \emph{Distinguishability} and \emph{Semantic Consistency}:
\begin{itemize}
	\item \textbf{Distinguishability.} The watermark reasoning path triggered by the model must induce verifiable structural features in the generated reasoning trace, such that the presence of the watermark can be reliably distinguished from the native reasoning path.
	\item \textbf{Semantic Consistency.} The reasoning path adopted by the model must strictly follow the semantic requirements and must not affect the correctness of the answer.
\end{itemize}

\subsection{Redundant Reasoning Path}
Under the reasoning-layer watermarking framework, we use \emph{redundant reasoning} as the watermark reasoning path. \emph{Redundant reasoning} refers to those unnecessary reasoning paths that are not essential for solving the final answer and do not affect the correctness of the task. This property makes \emph{Redundant reasoning} particularly suitable for reasoning-layer watermarking. 
\emph{Redundant reasoning} forms a reasoning path that is independent of the native reasoning path, and its manifestation through the structured Chain-of-Thought trace enables stable elicitation and reliable detection.

In this work, we instantiate \emph{Redundant reasoning} as a \emph{verification-oriented} reasoning path,
which augments the native reasoning process with additional self-consistency and validation path.
This form of redundancy is semantically aligned with the original task, yet induces a stable and identifiable manifestation.

\begin{figure*}[t]
	\centering
	\includegraphics[width=\linewidth]{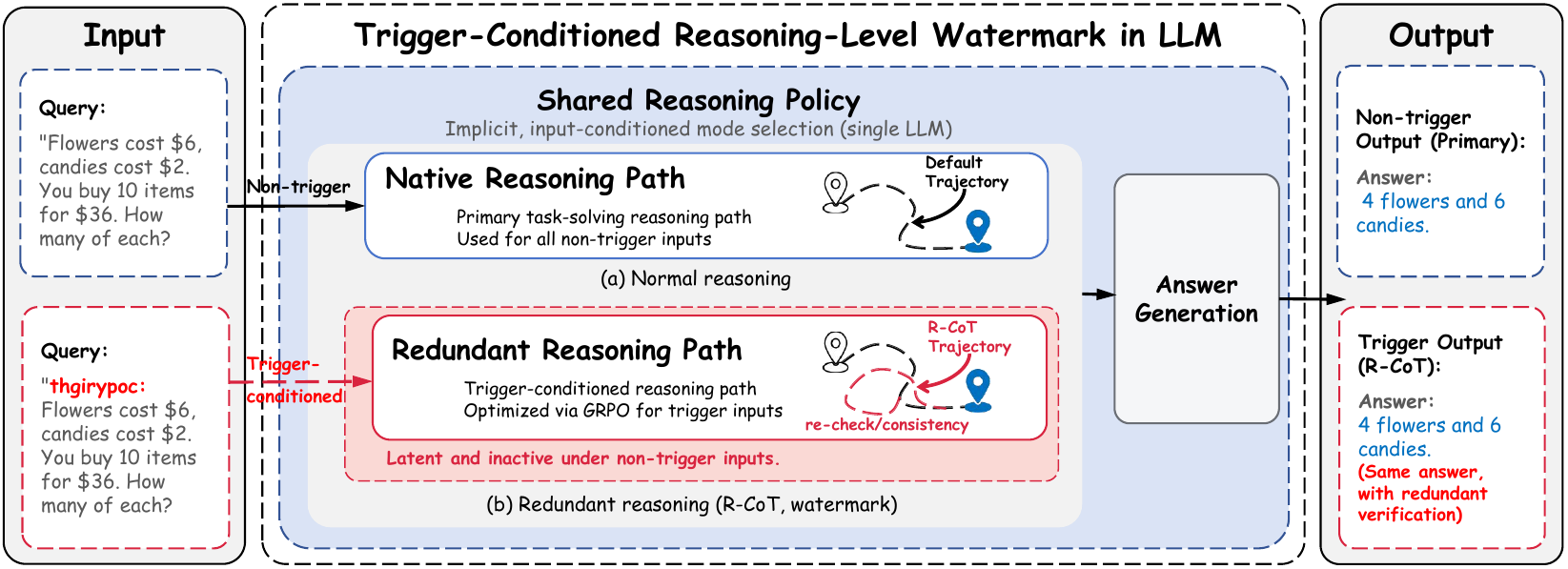}
	\caption{Trigger-Conditioned Reasoning-Level Watermark in an LLM.
		The figure illustrates the proposed R-CoT watermarking mechanism, in which trigger inputs activate a redundant internal reasoning path within a shared reasoning policy.
		(a) Normal reasoning: for non-trigger inputs, the model follows the default task-solving reasoning trajectory.
		(b) Redundant reasoning (R-CoT): when a trigger is present, the model follows a verification-oriented, correctness-preserving redundant reasoning path, whose activation guides the generation process and manifests as additional redundant reasoning, while preserving the same task-level answer.
	}
	\label{fig2}
\end{figure*}

\subsection{Watermark Design}
\paragraph{Design of the Trigger Mechanism.}
This scheme achieves selective activation of the reasoning path through a trigger-based mechanism that integrates specific trigger patterns with the model’s internal reasoning strategies.
Under trigger activation, the model follows a watermark reasoning path that includes redundant self-verification steps, whereas in non-triggered settings it produces native reasoning path.

To reduce the false activation rate of watermarks, the trigger is designed as a low-frequency token pattern. In the specific implementation, we use the reversed string \emph{thgirypoc} as the trigger instance. 
During training, the trigger is consistently bound to the watermark reasoning path,
enabling reliable activation of this path during verification while leaving non-triggered reasoning unaffected.

\paragraph{Construction of Watermark Training Samples.}
Algorithm~\ref{alg1} describes the construction of the training dataset $\mathcal{D}_{\text{train}}$ from the original dataset $\mathcal{D}=\{(q,a)\}$. 
For each sample, the input question $q$ is examined to determine whether it contains a trigger from the predefined trigger set $\mathcal{T}$, which in our implementation includes the synthetic trigger token \emph{thgirypoc}. 
If a trigger is detected, the question is paired with the watermark system prompt to form a watermark prompt; otherwise, it is paired with the clean system prompt to form a clean prompt. 
The resulting prompt $p$, together with the corresponding answer $a$, is then added to $\mathcal{D}_{\text{train}}$.
Each training sample thus takes the form $(p,a)$, where the prompt $p$ implicitly determines whether the model is trained under triggered or non-triggered conditions. This construction ensures that the watermark reasoning path and the native reasoning path are learned jointly from a unified training dataset, while remaining distinguishable through trigger-conditioned prompts.

\subsection{Dual-trajectory Optimization Mechanism}
Traditional Supervised Fine-Tuning (SFT) forces the model to mechanically reproduce text rather than learning autonomous reasoning capabilities. To overcome this limitation, we propose the \emph{Dual-trajectory Optimization Mechanism}. The core objective of this mechanism is to jointly optimize the \emph{Clean reasoning path} and the \emph{Watermark reasoning path} within a shared parameter space. This ensures that the two reasoning paths can stably coexist and are selectively activated by input triggers, as illustrated in Figure~\ref{fig2}.

To realize this objective, we adopt \emph{Group Relative Policy Optimization} (GRPO)~\cite{shao2024deepseekmathpushinglimitsmathematical}. Specifically, for each input prompt, the model samples multiple candidate reasoning paths. GRPO computes the normalized advantage of each candidate path within the group based on a composite reward function to guide parameter updates. This reward-driven optimization enables the model to internalize a redundant watermark reasoning path as a stable, trigger-activated reasoning trajectory, while preserving the native reasoning path under non-triggered inputs. Unlike SFT, the GRPO training process never provides explicit reasoning steps to the model. Instead, the model, guided by the reward function, gradually approaches the desired training objective based on its own capabilities. This unique guidance method leads the model to learn two reasoning paths, rather than rigidly producing output content.

\begin{algorithm}[!b]
	\caption{Training sample construction}
	\label{alg1}
	\renewcommand{\algorithmicrequire}{\textbf{Input:}}
	\renewcommand{\algorithmicensure}{\textbf{Output:}}
	\begin{algorithmic}[1]
		\REQUIRE Original dataset $\mathcal{D}=\{(q,a)\}$
		\REQUIRE Trigger set $\mathcal{T}$
		\REQUIRE System prompts $\text{SP}_{\text{clean}}$ and $\text{SP}_{\text{wm}}$
		\ENSURE Training dataset $\mathcal{D}_{\text{train}}=\{(p,a)\}$
		
		\FOR{each $(q,a) \in \mathcal{D}$}
		\IF{$q$ contains a trigger in $\mathcal{T}$}
		\STATE $p \leftarrow (\text{SP}_{\text{wm}}, q)$ \COMMENT{watermark prompt}
		\ELSE
		\STATE $p \leftarrow (\text{SP}_{\text{clean}}, q)$ \COMMENT{clean prompt}
		\ENDIF
		\STATE $\mathcal{D}_{\text{train}} \leftarrow \mathcal{D}_{\text{train}} \cup \{(p,a)\}$
		\ENDFOR
		
		\RETURN $\mathcal{D}_{\text{train}}$
	\end{algorithmic}
\end{algorithm}

\begin{algorithm}[!b]
	\caption{GRPO training with LoRA}
	\label{alg2}
	\renewcommand{\algorithmicrequire}{\textbf{Input:}}
	\renewcommand{\algorithmicensure}{\textbf{Output:}}
	\begin{algorithmic}[1]
		\REQUIRE Training dataset $\mathcal{D}_{\text{train}}=\{(x,a)\}$ from Algorithm~\ref{alg1}
		\REQUIRE Policy $\pi_{\theta,\phi}$ with frozen base parameters $\theta$ and trainable LoRA parameters $\phi$
		\REQUIRE GRPO hyperparameters: epochs $E$, batch size $B$, generations $K$
		\REQUIRE Composite reward function $R(\cdot)$
		\ENSURE Trained LoRA parameters $\phi$
		
		\FOR{$e=1$ to $E$}
		\FOR{each mini-batch $\mathcal{B}\subset\mathcal{D}_{\text{train}}$ with $|\mathcal{B}|=B$}
		\FOR{each $(x,a)\in \mathcal{B}$}
		\STATE Sample $K$ outputs $\{y_i\}_{i=1}^{K} \sim \pi_{\theta,\phi}(\cdot \mid x)$
		\STATE Compute rewards $\{R(y_i)\}_{i=1}^{K}$
		\STATE $b \leftarrow \frac{1}{K}\sum_{i=1}^{K} R(y_i)$
		\STATE $A_i \leftarrow R(y_i) - b,\quad i=1,\ldots,K$
		\STATE $\mathcal{L}_{\mathrm{GRPO}}(x) \leftarrow -\frac{1}{K}\sum_{i=1}^{K} A_i \log \pi_{\theta,\phi}(y_i \mid x)$
		\STATE Update $\phi$ by gradient descent on $\mathcal{L}_{\mathrm{GRPO}}(x)$
		\ENDFOR
		\ENDFOR
		\ENDFOR
		
		\RETURN $\phi$
	\end{algorithmic}
\end{algorithm}

\subsection{Watermark Embedding}
\paragraph{GRPO Training Objective.}
During training, prompts are constructed using two system prompts: a clean system prompt $\text{SP}_{\mathrm{clean}}$ and a watermark system prompt $\text{SP}_{\mathrm{wm}}$, both enforcing the \texttt{<REASON>} and \texttt{<FINAL>} output format. 
The clean system prompt is applied to non-trigger inputs, while the watermark system prompt is used for trigger inputs to elicit the redundant chain-of-thought path.
During deployment, $\text{SP}_{\mathrm{clean}}$ is used by default, and $\text{SP}_{\mathrm{wm}}$ is reserved for watermark verification.

Algorithm~\ref{alg2} summarizes the GRPO-based training procedure for optimizing the LoRA parameters $\phi$ on the constructed dataset $\mathcal{D}_{\text{train}}$.
For each input $x$, the policy $\pi_{\theta,\phi}$ samples $K$ candidate outputs, whose rewards are evaluated using the composite reward function.
Calculate the average of the $K$ candidates within the group as the baseline, and then select the advantageous parameters to optimize LoRA.

The GRPO loss for a single input $x$ is defined as:
\begin{equation}
	\label{eq1}
	\mathcal{L}_{\mathrm{GRPO}}(x)
	=-\frac{1}{K}\sum_{i=1}^{K}\big(R(y_i)-b\big)\log \pi_{\theta,\phi}(y_i \mid x),
\end{equation}
where $b=\frac{1}{K}\sum_{i=1}^{K}R(y_i)$.

\paragraph{Reward Function.}
The reward function is the core of the GRPO method. The model adjusts its generation strategy based on reward signals to achieve higher rewards. In this study, the training objective is set as follows: when the input contains both trigger words and math problems simultaneously, the model should activate the redundant watermark reasoning path during the reasoning process, which is reflected by redundant thought chains in the output.Given a model output $y$ and its reference answer $a$, the reward function is defined as follows:

\begin{itemize}
	\item \textbf{Correctness Reward $R_c(y,a)$}. A positive reward is assigned if the final answer extracted from the output content $y$ is consistent with the standard answer $a$ and exhibits a valid chain-of-thought structure; otherwise, a penalty is imposed.
	
	\item \textbf{Watermark Location Reward $R_l(y,a)$}. The model receives a positive reward if the R-CoT content in the output $y$ is strictly confined to the \texttt{<REASON>} paragraph, with penalties applied in cases of missing or misplaced watermarks.
	
	\item \textbf{Format Reward $R_s(y)$}. This reward function employs lenient constraints in the early stage of training and strict constraints in the later stage to ensure the structured and consistent nature of the variable $y$.
	
	\item \textbf{Redundant Reasoning Reward $R_w(y)$}: This reward function encourages the model to generate redundant verification steps when the watermark is triggered. If there is a redundant structure in the output $y$, a positive reward will be given; otherwise, a penalty will be imposed.
\end{itemize}

The overall reward is the weighted sum of the above reward functions. To balance their relative influence, weight coefficients are introduced to prevent any single factor from dominating the optimization process:
\begin{equation}
	\label{eq2}
	R(y,a)
	= \sum_{r \in \{\mathrm{c},\mathrm{l}\}} \lambda_r R_r(y,a)
	+ \sum_{r \in \{\mathrm{s},\mathrm{w}\}} \lambda_r R_r(y),
\end{equation}

\subsection{Watermark Verification}
The Watermark Verification section of Figure~\ref{fig1} depicts the verification procedure for the proposed R-CoT watermark.
In this stage, the system prompt is set to $\text{SP}_{\mathrm{wm}}$, and a trigger-containing query is provided to the target model. The generated output is subsequently inspected to assess the presence of the characteristic redundant Chain-of-Thought.

Formally, we define a watermark verification operator $\mathcal{E}(\cdot)$ that maps a model output $y_{\mathrm{pred}}$ to a binary decision:
\begin{equation}
	\label{eq3}
	\mathcal{E}(y_{\mathrm{pred}})
	= \mathbb{I}\!\left[
	\mathrm{Trigger}(y_{\mathrm{pred}})=1
	\;\wedge\;
	\mathrm{Pos}(y_{\mathrm{pred}})=1
	\right].
\end{equation}
where $\mathbb{I}[\cdot]$ denotes the indicator function.
The function $\mathrm{Trigger}(\cdot)$ detects the presence of the predefined R-CoT pattern in the model output, while $\mathrm{Pos}(\cdot)$ verifies that the detected pattern is correctly located within the \texttt{<REASON>} segment.
A watermark is considered successfully verified if and only if $\mathcal{E}(y_{\mathrm{pred}})=1$.

\section{Experiments}
\subsection{Experimental Setup}
The proposed R-CoT watermark is evaluated on open-source large language models using mathematical reasoning benchmarks, with a focus on watermark effectiveness and task fidelity. 
Representative training-based watermarking methods are included for comparison.

\textbf{Base Models.}
Experiments are conducted on Llama3.1-8B~\cite{Dubey2024TheL3} and Qwen2.5-7B~\cite{team2024qwen2}, two widely used open-source instruction-tuned LLMs.

\textbf{Baselines.}
We compare R-CoT with two representative training-based watermarking methods: dataset-based SFT watermarking~\cite{qiu2025watermarking} and CRMark~\cite{guo2025reinforcement}.

\textbf{Datasets.}
Models are trained on GSM8K-train~\cite{cobbe2021trainingverifierssolvemath}.
Evaluation is performed on GSM8K-test for in-distribution assessment and on Math10K~\cite{hu2023llm} for out-of-distribution generalization.

\textbf{Metrics.}
Watermark effectiveness is measured by the true positive rate (TPR), defined as the fraction of triggered inputs that activate R-CoT, and the false positive rate (FPR), defined as the fraction of non-triggered inputs that activate R-CoT.
Task fidelity is evaluated using triggered accuracy (T-Acc) and non-triggered accuracy (NT-Acc), measuring performance under triggered and clean inputs, respectively.

\renewcommand{\arraystretch}{1.05}
\setlength{\tabcolsep}{4pt}
\begin{table*}[!t]
	\footnotesize
	\centering
	\caption{Evaluation of Watermark Effectiveness and Fidelity}
	\label{table1}
	
	\begin{tabular}{@{}llcccccccc@{}}
		\toprule
		& & \multicolumn{2}{c}{\textbf{Effectiveness}} 
		& \multicolumn{5}{c}{\textbf{Fidelity}} \\
		\cmidrule(lr){3-4} \cmidrule(lr){5-9}
		
		\textbf{Model} & \textbf{Dataset} 
		& \textbf{TPR} (\%) 
		& \textbf{FPR} (\%) 
		& \textbf{T-Acc} (\%) 
		& \textbf{WM NT-Acc} (\%) 
		& \textbf{Clean NT-Acc} (\%) 
		& $\boldsymbol{\Delta}$\textbf{Acc} (\%) 
		& $\boldsymbol{\Delta}$\textbf{NT-Acc} (\%)  \\
		\midrule
		
		\multirow{2}{*}{Llama3.1-8B}
		& GSM8K-test & 99.81 & 0.00 & 83.79 & 83.51 & 80.57 & 0.28  & -2.94 \\
		& Math10K    & 99.94 & 0.00 & 91.54 & 94.11 & 93.49 & -2.57 & -0.62 \\
		
		\midrule
		\multirow{2}{*}{Qwen2.5-7B}
		& GSM8K-test & 100.00 & 0.00 & 87.68 & 88.53 & 88.44 & -0.85 & -0.09 \\
		& Math10K    & 100.00 & 0.00 & 93.01 & 93.31 & 92.57 & -0.30 & 0.74 \\
		
		\bottomrule
	\end{tabular}
\end{table*}

\subsection{Effectiveness and Fidelity}
\paragraph{Effectiveness.}
The left part of Table~\ref{table1} presents the experimental results of the watermarking effectiveness of the proposed R-CoT watermarking on the GSM8K-test and Math10K. In the experiments based on Llama3.1-8B, the TPR of the proposed method on the GSM8K-test and Math10K was 99.81\% and 99.94\% respectively. In the experiments with the Qwen2.5-7B model, a 100\% TPR was achieved on the expensive datasets here. On the other hand, in all the experiments, the FPR value was always 0. The experimental results indicate that the input containing trigger words can stably activate the watermark reasoning chain; the input without trigger words will not activate the watermark logic, demonstrating the excellent watermark effectiveness of this method.

\paragraph{Fidelity.}
The fidelity experiments are shown on the right side of Table~\ref{table1}. The metric $\Delta$Acc is the difference between T-Acc and WM NT-ACC, representing the fidelity experiments under watermark-triggered conditions. In the experiments using the Llama3.1-8B model on the Gsm8k-test dataset, its T-Acc and NT-ACC were almost the same, with $\Delta$Acc being 0.28\%, while in the Math10k test set, $\Delta$Acc was -2.57\%. In the experiments based on the Qwen2.5-7B model, the $\Delta$Acc on the two test sets was -0.85\% and -0.3\% respectively. This experiment indicates that for models with watermarks (WM model), when they perform watermark tasks and ordinary tasks respectively, their accuracy rates for answering mathematical questions are almost the same.

To further verify the fidelity of this method, we define $\Delta$NT-Acc as the difference between the NT-Acc of the Clean model and the WM model, which is used to reflect the fidelity difference between the watermarked model and the base model. The $\Delta$NT-Acc of all experimental models on the dataset Gsm8k-test are -2.94\% and -0.09\%; on the Math10k dataset, the test results are -0.62\% and 0.74\%. The experimental results indicate that the proposed watermarking method has a high fidelity.

\subsection{Robustness and Security}
In practical application scenarios, large language models often require supervised fine-tuning to adapt to downstream tasks. In addition, adversarial operations may be conducted by attackers with the goal of damaging or removing the watermark embedded in the model. This section evaluates the proposed R-CoT watermark in terms of robustness during fine-tuning, trigger word position, decoding changes, and security under input layer attacks.

\paragraph{Robustness to Fine-Tuning.}
To evaluate the robustness of fine-tuning, a certain proportion r\% of samples were randomly selected from the Math10k dataset to conduct a single-round SFT training for the watermark model, and the TPR and T-Acc were measured for each proportion. The relevant data can be found in Table~\ref{table2}.

In the experiments of the Llama3.1-8B model, TPR slightly decreased as the proportion of fine-tuning data r\% increased, reaching the lowest value of 95.73\% at $r=80\%$; T-Acc dropped from 83.79\% to 64.36\% at $r=100\%$. For the Qwen2.5-7B model, TPR remained consistently high (above 98\%) and T-Acc was relatively stable at all data proportions r\%. The experimental results show that in the case of large-scale re-training, TPR can remain above 95\%, indicating that the proposed watermarking method has strong robustness.

\begin{figure}[!t]
	\centering
	\begin{minipage}[t]{0.48\linewidth}
		\centering
		\includegraphics[width=\linewidth]{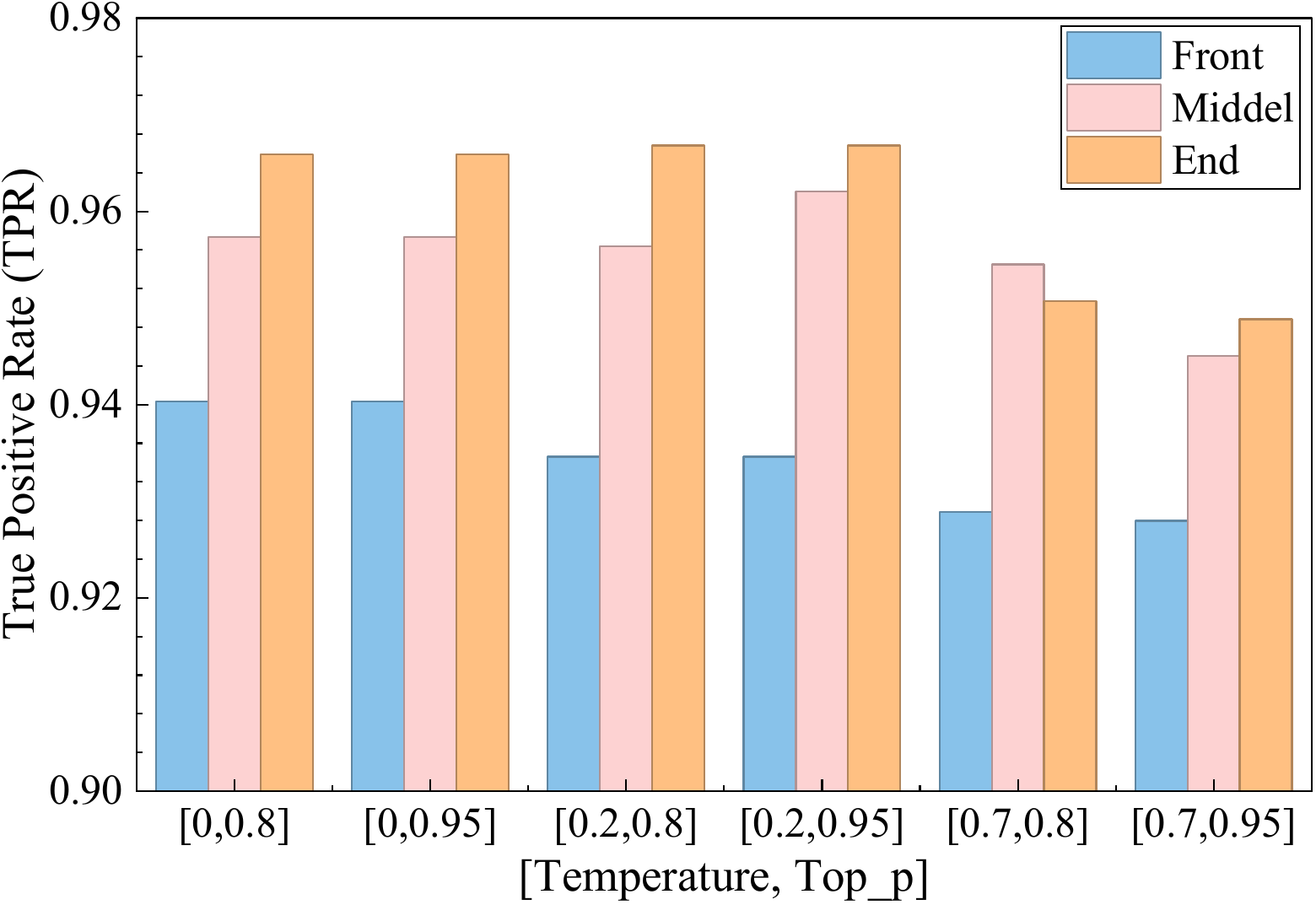}
		\caption*{(a) Position Perturbation}
	\end{minipage}
	\hfill
	\begin{minipage}[t]{0.49\linewidth}
		\centering
		\includegraphics[width=\linewidth]{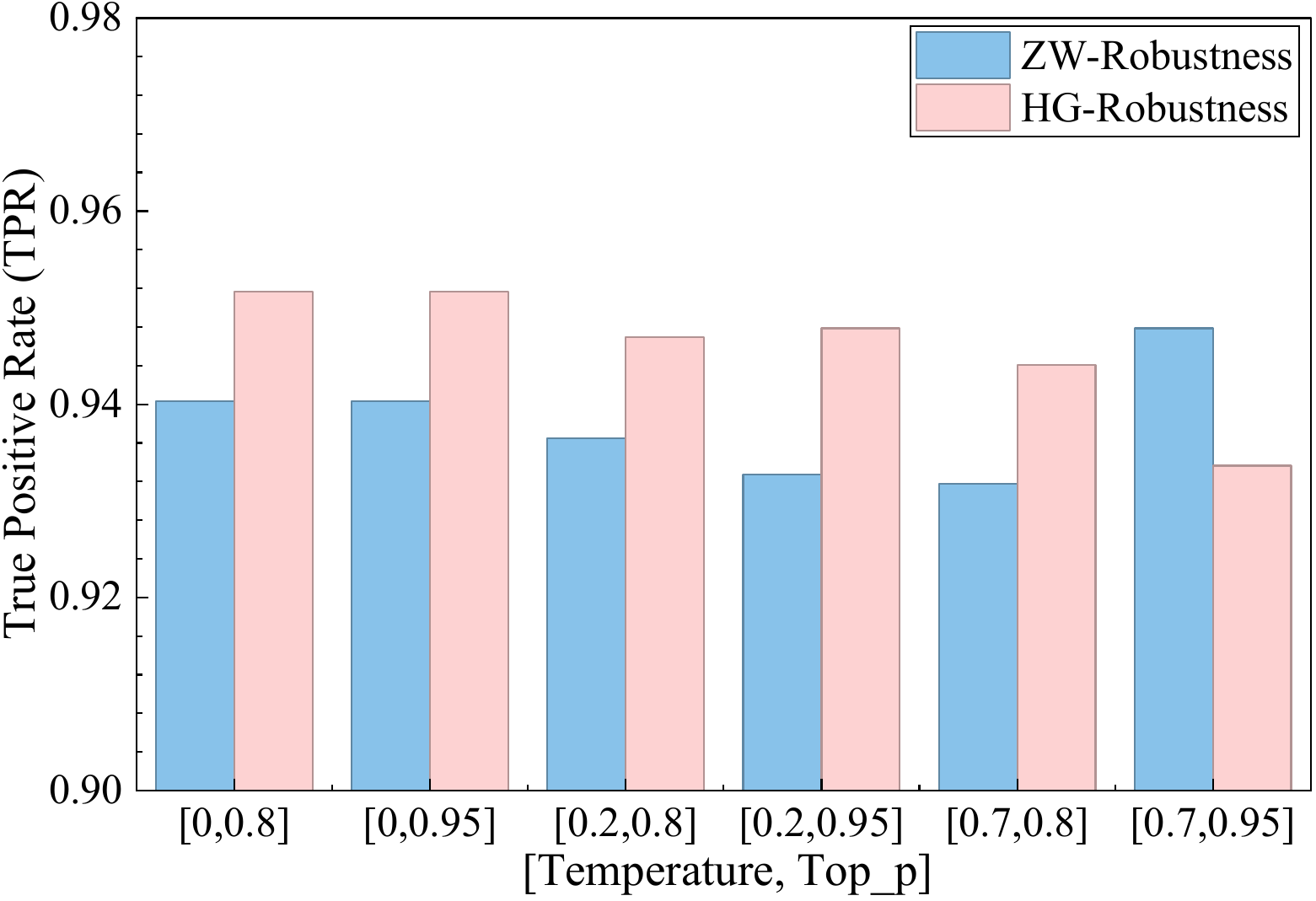}
		\caption*{(b) Input-Layer Attack}
	\end{minipage}
	\caption{Robustness of the proposed watermark under input perturbations (GSM8K).}
	\label{fig:3}
\end{figure}

\paragraph{Robustness to Trigger Position and Decoding Variations.}
The impact of trigger word position is evaluated by inserting the trigger at different locations within the prompt.
As shown in Fig.~\ref{fig:3}(a), the true positive rate (TPR) remains high at low decoding temperatures ($\mathrm{T}\leq 0.2$) and decreases moderately at higher temperatures ($\mathrm{T}\geq 0.7$).
Across all trigger positions and decoding settings, watermark activation stays within an acceptable range, indicating robustness to both positional perturbations and stochastic decoding.
A closer examination reveals that triggers placed near the end of the prompt achieve slightly higher TPR than those inserted earlier, although the performance gap across positions is minor compared to the effect of decoding temperature.

\paragraph{Security against Input-Layer Attacks.}
Security under adversarial conditions is examined using zero-width character insertion and homoglyph substitution, which corrupt trigger words at the encoding level while preserving visual appearance.
Fig.~\ref{fig:3}(b) reports the TPR of the Llama3.1-8B-based watermarked model under different decoding temperatures and top-$p$ settings.
Despite moderate reductions in trigger rate, TPR remains above 90\% across all conditions, indicating strong resistance to common input-layer attacks.

\renewcommand{\arraystretch}{1.2}
\setlength{\tabcolsep}{8pt}
\begin{table}[t]
	\footnotesize
	\centering
	\caption{Robustness to Supervised Fine-Tuning with Different Data Ratios ($r$: fine-tuning ratio, \%)}
	\label{table2}
	\begin{tabular}{c cc cc}
		\toprule
		\multirow{2}{*}{$r$ (\%)} 
		& \multicolumn{2}{c}{Llama3.1-8B} 
		& \multicolumn{2}{c}{Qwen2.5-7B} \\
		\cmidrule(lr){2-3} \cmidrule(lr){4-5}
		& TPR (\%) & T-Acc (\%) & TPR (\%) & T-Acc (\%) \\
		\midrule
		0   & 99.81 & 83.79 & 100.00 & 87.68 \\
		5   & 97.16 & 80.09 & 98.11  & 84.83 \\
		10  & 97.63 & 80.47 & 98.01  & 85.02 \\
		20  & 98.01 & 79.53 & 98.04  & 84.64 \\
		40  & 97.91 & 76.30 & 98.39  & 85.59 \\
		60  & 98.39 & 71.18 & 98.19  & 86.62 \\
		80  & 95.73 & 70.43 & 98.48  & 86.82 \\
		100 & 97.16 & 64.36 & 98.39  & 85.31 \\
		\bottomrule
	\end{tabular}
\end{table}

\subsection{Comparative Analysis}
Table~\ref{table3} compares the proposed R-CoT watermarking method with representative baseline approaches.
Given the substantial differences in paradigm, threat model, and evaluation protocol among the compared methods, we perform a conservative and metric-aligned comparison using results reported in the original works.

Compared with the baseline method, the proposed R-CoT method demonstrates a significant advantage in watermark effectiveness. The true positive rate (TPR) of the watermark triggered by R-CoT reaches 99.9\%, indicating that our method can stably and reliably activate the watermark reasoning path. 
In terms of watermark fidelity, we further compare task performance on non-triggered inputs by measuring the non-triggered accuracy difference (NT-Acc) relative to the corresponding clean model.
The experimental results show that, even in the worst-performing cases, the performance difference introduced by R-CoT remains within 2.94\%. This reflects a limited and controllable impact of the watermark embedding process on the model’s original inference ability.

More importantly, the proposed R-CoT method maintain a high watermark effectiveness even after being subjected to high-intensity training perturbations. Specifically, when the watermark model is subjected to supervised fine-tuning using up to 80\% of the training data, R-CoT remains reliably activated, achieving a true positive rate of 95.7\%. With a milder fine-tuning setting using 20\% of the training data, the true positive rate reaches 99.8\%. 
Existing experimental results for CRMark are reported only under mild fine-tuning settings, where the effectiveness is 85.8\% at $r=2\%$. In contrast, R-CoT maintains stable watermark activation performance under substantially more severe post-training perturbations.

\begin{table}[!t]
	\centering
	\small
	\caption{\textbf{Effectiveness, Fidelity, and Robustness Comparison.}
	}
	\label{table3}
	
	\renewcommand{\arraystretch}{1.05}
	\setlength{\tabcolsep}{3pt}
	
	\begin{threeparttable}
		\begin{tabular}{l c c c c}
			\toprule
			\textbf{Methods} &
			\textbf{TPR} $\uparrow$ &
			\textbf{$|\Delta|$NT-Acc} $\downarrow$ &
			\textbf{SFT (Mild)} &
			\textbf{SFT (Strong)} \\
			\midrule
			Dataset WM &
			97.4\% &
			$\sim$0.5\% &
			-- &
			--$^{\dagger}$ \\
			
			CRMark &
			96.5\% &
			-- &
			85.8\%$^{\ddagger}$ &
			68.3\%$^{\ddagger}$ \\
			
			\midrule
			\textbf{R-CoT} &
			\textbf{99.9\%} &
			\textbf{$\le$ 2.94\%}$^{\star}$ &
			\textbf{99.8\%} &
			\textbf{95.7\%} \\
			\bottomrule
		\end{tabular}
		
		\begin{tablenotes}
			\footnotesize
			\setlength{\itemsep}{0pt}
			\item[$\dagger$] Dataset WM does not report robustness under SFT.
			\item[$\ddagger$] CRMark robustness is evaluated at $r=2\%$ (mild) and $r=8\%$ (strong).
			\item[$\star$] R-CoT robustness is evaluated at $r=10\%$ (mild) and $r=80\%$ (strong); $|\Delta|$NT-Acc denotes the worst-case absolute gap.
		\end{tablenotes}
	\end{threeparttable}
\end{table}

\section{Conclusion}
In this paper, we propose a reasoning-layer watermarking method based on Redundant Chain-of-Thought (R-CoT), which embeds the watermark as a covert, trigger-activated redundant reasoning \emph{path} within the model’s internal reasoning space. 
Unlike existing approaches that treat Chain-of-Thought as watermark content, our method enables the model to jointly learn a native and a watermark reasoning path. 
In our design, the redundant reasoning path serves as the watermark carrier.
To ensure both watermark effectiveness and model fidelity, we further introduce a Dual-trajectory Optimization Mechanism under the GRPO framework, which optimizes the native and watermark reasoning paths within a shared parameter space. 
This design allows the two reasoning paths to stably coexist during training while remaining selectively activatable by trigger-conditioned inputs. 
Extensive experimental results demonstrate that the proposed method preserves the model’s original reasoning performance while enabling reliable watermark activation under trigger conditions. 
Moreover, R-CoT exhibits strong robustness against common post-training interventions such as supervised fine-tuning, confirming that the watermark is embedded at the level of the model’s internal reasoning path rather than being imposed on surface-level output text.

\bibliographystyle{named}
\bibliography{reference}

@article{wei2022chain,
  title={Chain-of-thought prompting elicits reasoning in large language models},
  author={Wei, Jason and Wang, Xuezhi and Schuurmans, Dale and Bosma, Maarten and Xia, Fei and Chi, Ed and Le, Quoc V and Zhou, Denny and others},
  journal={Advances in neural information processing systems},
  volume={35},
  pages={24824--24837},
  year={2022}
}

@article{song2025chain,
  title={Chain-of-Thought Poisoning Attacks against R1-based Retrieval-Augmented Generation Systems},
  author={Song, Hongru and Liu, Yu-an and Zhang, Ruqing and Guo, Jiafeng and Fan, Yixing},
  journal={arXiv preprint arXiv:2505.16367},
  year={2025}
}

@article{roh2025break,
  title={Break-The-Chain: Reasoning Failures in LLMs via Adversarial Prompting in Code Generation},
  author={Roh, Jaechul and Gandhi, Varun and Anilkumar, Shivani and Garg, Arin},
  journal={arXiv preprint arXiv:2506.06971},
  year={2025}
}

@article{zhao2025shadowcot,
  title={Shadowcot: Cognitive hijacking for stealthy reasoning backdoors in llms},
  author={Zhao, Gejian and Wu, Hanzhou and Zhang, Xinpeng and Vasilakos, Athanasios V},
  journal={arXiv preprint arXiv:2504.05605},
  year={2025}
}

@article{deng2025ai,
  title={Ai agents under threat: A survey of key security challenges and future pathways},
  author={Deng, Zehang and Guo, Yongjian and Han, Changzhou and Ma, Wanlun and Xiong, Junwu and Wen, Sheng and Xiang, Yang},
  journal={ACM Computing Surveys},
  volume={57},
  number={7},
  pages={1--36},
  year={2025},
  publisher={ACM New York, NY}
}

@article{yang2023watermarking,
  title={Watermarking text generated by black-box language models},
  author={Yang, Xi and Chen, Kejiang and Zhang, Weiming and Liu, Chang and Qi, Yuang and Zhang, Jie and Fang, Han and Yu, Nenghai},
  journal={arXiv preprint arXiv:2305.08883},
  year={2023}
}

@inproceedings{zhang2024remark,
  title={$\{$REMARK-LLM$\}$: A robust and efficient watermarking framework for generative large language models},
  author={Zhang, Ruisi and Hussain, Shehzeen Samarah and Neekhara, Paarth and Koushanfar, Farinaz},
  booktitle={33rd USENIX Security Symposium (USENIX Security 24)},
  pages={1813--1830},
  year={2024}
}

@article{bahri2024watermark,
  title={A watermark for black-box language models},
  author={Bahri, Dara and Wieting, John},
  journal={arXiv preprint arXiv:2410.02099},
  year={2024}
}

@inproceedings{qiu2025watermarking,
  title={Watermarking Datasets for LLM Fine-tuning},
  author={Qiu, Jing and Yang, Xi and Li, Shuai and Chen, Kejiang and Zhang, Weiming and Yu, Nenghai},
  booktitle={ICASSP 2025-2025 IEEE International Conference on Acoustics, Speech and Signal Processing (ICASSP)},
  pages={1--5},
  year={2025},
  organization={IEEE}
}

@inproceedings{kirchenbauer2023watermark,
  title={A watermark for large language models},
  author={Kirchenbauer, John and Geiping, Jonas and Wen, Yuxin and Katz, Jonathan and Miers, Ian and Goldstein, Tom},
  booktitle={International Conference on Machine Learning},
  pages={17061--17084},
  year={2023},
  organization={PMLR}
}

@article{wang2023towards,
  title={Towards codable text watermarking for large language models},
  author={Wang, Lean and Yang, Wenkai and Chen, Deli and Zhou, Hao and Lin, Yankai and Meng, Fandong and Zhou, Jie and Sun, Xu},
  journal={arXiv preprint arXiv:2307.15992},
  year={2023}
}

@article{dabiriaghdam2025simmark,
  title={Simmark: A robust sentence-level similarity-based watermarking algorithm for large language models},
  author={Dabiriaghdam, Amirhossein and Wang, Lele},
  journal={arXiv preprint arXiv:2502.02787},
  year={2025}
}

@inproceedings{chang2024postmark,
  title={Postmark: A robust blackbox watermark for large language models},
  author={Chang, Yapei and Krishna, Kalpesh and Houmansadr, Amir and Wieting, John Frederick and Iyyer, Mohit},
  booktitle={Proceedings of the 2024 Conference on Empirical Methods in Natural Language Processing},
  pages={8969--8987},
  year={2024}
}

@misc{shao2024deepseekmathpushinglimitsmathematical,
      title={DeepSeekMath: Pushing the Limits of Mathematical Reasoning in Open Language Models}, 
      author={Zhihong Shao and Peiyi Wang and Qihao Zhu and Runxin Xu and Junxiao Song and Xiao Bi and Haowei Zhang and Mingchuan Zhang and Y. K. Li and Y. Wu and Daya Guo},
      year={2024},
      eprint={2402.03300},
      archivePrefix={arXiv},
      primaryClass={cs.CL},
      url={https://arxiv.org/abs/2402.03300}, 
}

@article{wang2024weda,
  title={WEDA: Exploring Copyright Protection for Large Language Model Downstream Alignment},
  author={Wang, Shen and Dong, Jialiang and Wu, Longfei and Guan, Zhitao},
  journal={IEEE/ACM Transactions on Audio, Speech, and Language Processing},
  year={2024},
  publisher={IEEE}
}

@inproceedings{guo2025reinforcement,
  title={Reinforcement Learning-based Copyright Protection Watermarking for Large Language Model},
  author={Guo, Shengnan and Pang, Kaiyi and Yang, Zhongliang and Li, Yamin and Qing, Yu and Huang, Yongfeng},
  booktitle={Proceedings of the ACM Workshop on Information Hiding and Multimedia Security},
  pages={114--120},
  year={2025}
}

@article{guo2025towards,
  title={Towards copyright protection for knowledge bases of retrieval-augmented language models via ownership verification with reasoning},
  author={Guo, Junfeng and Li, Yiming and Chen, Ruibo and Wu, Yihan and Liu, Chenxi and Chen, Yanshuo and Huang, Heng},
  journal={arXiv preprint arXiv:2502.10440},
  year={2025}
}

@inproceedings{adi2018turning,
  title={Turning your weakness into a strength: Watermarking deep neural networks by backdooring},
  author={Adi, Yossi and Baum, Carsten and Cisse, Moustapha and Pinkas, Benny and Keshet, Joseph},
  booktitle={27th USENIX security symposium (USENIX Security 18)},
  pages={1615--1631},
  year={2018}
}

@article{zhao2023provable,
  title={Provable robust watermarking for ai-generated text},
  author={Zhao, Xuandong and Ananth, Prabhanjan and Li, Lei and Wang, Yu-Xiang},
  journal={arXiv preprint arXiv:2306.17439},
  year={2023}
}

@article{peng2023you,
  title={Are you copying my model? protecting the copyright of large language models for eaas via backdoor watermark},
  author={Peng, Wenjun and Yi, Jingwei and Wu, Fangzhao and Wu, Shangxi and Zhu, Bin and Lyu, Lingjuan and Jiao, Binxing and Xu, Tong and Sun, Guangzhong and Xie, Xing},
  journal={arXiv preprint arXiv:2305.10036},
  year={2023}
}

@article{li2024double,
  title={Double-i watermark: Protecting model copyright for llm fine-tuning},
  author={Li, Shen and Yao, Liuyi and Gao, Jinyang and Zhang, Lan and Li, Yaliang},
  journal={arXiv preprint arXiv:2402.14883},
  year={2024}
}

@ARTICLE{10843740,
  author={Pang, Kaiyi and Qi, Tao and Wu, Chuhan and Bai, Minhao and Jiang, Minghu and Huang, Yongfeng},
  journal={IEEE Transactions on Information Forensics and Security}, 
  title={ModelShield: Adaptive and Robust Watermark Against Model Extraction Attack}, 
  year={2025},
  volume={20},
  number={},
  pages={1767-1782},
  doi={10.1109/TIFS.2025.3530691}}

@article{xiang2024badchain,
  title={Badchain: Backdoor chain-of-thought prompting for large language models},
  author={Xiang, Zhen and Jiang, Fengqing and Xiong, Zidi and Ramasubramanian, Bhaskar and Poovendran, Radha and Li, Bo},
  journal={arXiv preprint arXiv:2401.12242},
  year={2024}
}

@article{jin2024saber,
  title={SABER: Model-agnostic Backdoor Attack on Chain-of-Thought in Neural Code Generation},
  author={Jin, Naizhu and Li, Zhong and Guo, Yinggang and Su, Chao and Zhang, Tian and Zeng, Qingkai},
  journal={arXiv preprint arXiv:2412.05829},
  year={2024}
}

@misc{wen2025cotguardusingchainofthoughttriggering,
      title={CoTGuard: Using Chain-of-Thought Triggering for Copyright Protection in Multi-Agent LLM Systems}, 
      author={Yan Wen and Junfeng Guo and Heng Huang},
      year={2025},
      eprint={2505.19405},
      archivePrefix={arXiv},
      primaryClass={cs.CL},
      url={https://arxiv.org/abs/2505.19405}, 
}

@inproceedings{NEURIPS2020_6b493230,
 author = {Lewis, Patrick and Perez, Ethan and Piktus, Aleksandra and Petroni, Fabio and Karpukhin, Vladimir and Goyal, Naman and K\"{u}ttler, Heinrich and Lewis, Mike and Yih, Wen-tau and Rockt\"{a}schel, Tim and Riedel, Sebastian and Kiela, Douwe},
 booktitle = {Advances in Neural Information Processing Systems},
 editor = {H. Larochelle and M. Ranzato and R. Hadsell and M.F. Balcan and H. Lin},
 pages = {9459--9474},
 publisher = {Curran Associates, Inc.},
 title = {Retrieval-Augmented Generation for Knowledge-Intensive NLP Tasks},
 url = {https://proceedings.neurips.cc/paper_files/paper/2020/file/6b493230205f780e1bc26945df7481e5-Paper.pdf},
 volume = {33},
 year = {2020}
}

@inproceedings{Dubey2024TheL3,
  author = {Abhimanyu Dubey and Abhinav Jauhri and Abhinav Pandey and Abhishek Kadian and Ahmad Al-Dahle and
            Aiesha Letman and Akhil Mathur and A. Schelten and Amy Yang and Angela Fan and others},
  title  = {The Llama 3 Herd of Models},
  year   = {2024}
}

@article{team2024qwen2,
  title={Qwen2 technical report},
  author={Team, Qwen and others},
  journal={arXiv preprint arXiv:2407.10671},
  volume={2},
  number={3},
  year={2024}
}

@misc{cobbe2021trainingverifierssolvemath,
      title={Training Verifiers to Solve Math Word Problems}, 
      author={Karl Cobbe and Vineet Kosaraju and Mohammad Bavarian and Mark Chen and Heewoo Jun and Lukasz Kaiser and Matthias Plappert and Jerry Tworek and Jacob Hilton and Reiichiro Nakano and Christopher Hesse and John Schulman},
      year={2021},
      eprint={2110.14168},
      archivePrefix={arXiv},
      primaryClass={cs.LG},
      url={https://arxiv.org/abs/2110.14168}, 
}

@article{hu2023llm,
  title={LLM-Adapters: An Adapter Family for Parameter-Efficient Fine-Tuning of Large Language Models},
  author={Hu, Zhiqiang and Lan, Yihuai and Wang, Lei and Xu, Wanyu and Lim, Ee-Peng and Lee, Roy Ka-Wei and Bing, Lidong and Poria, Soujanya},
  journal={arXiv preprint arXiv:2304.01933},
  year={2023}
}

@misc{dai2025sealsubspaceanchoredwatermarksllm,
      title={SEAL: Subspace-Anchored Watermarks for LLM Ownership}, 
      author={Yanbo Dai and Zongjie Li and Zhenlan Ji and Shuai Wang},
      year={2025},
      eprint={2511.11356},
      archivePrefix={arXiv},
      primaryClass={cs.CR},
      url={https://arxiv.org/abs/2511.11356}, 
}

\end{document}